# Spin-transport in an organic semiconductor without free charge carrier involvement

H. Popli,[1] J. Wang,[1] X. Liu,[1] E. Lafalce,[1] T. H. Tennahewa,[1] H. Malissa,[1] Z. V. Vardeny,[1] and C. Boehme[1]

[1]*Department of Physics and Astronomy, University of Utah, Salt Lake City, Utah 84112, USA*

## Abstract

We have experimentally tested the hypothesis of free charge carrier mediated spin-transport in the small molecule organic semiconductor $Alq_3$ at room temperature. A spin current was pumped into this material by pulsed ferromagnetic resonance of an adjacent NiFe layer, while a charge current resulting from this spin current via the inverse spin-Hall effect (ISHE) was detected in a Pt layer adjacent on the other side of the $Alq_3$ layer, confirming a pure spin current through the $Alq_3$ layer. Charge carrier spin states in $Alq_3$, were then randomized by simultaneous application of electron paramagnetic resonance (EPR). No influence of the EPR excitation on the ISHE current was found, implying that spin-transport is not mediated by free charge-carriers in $Alq_3$.

It has been a long-standing open question whether the fundamental physical nature of spin transport and charge transport in organic semiconductor materials occurs via the same electronic states and mechanisms [1-3], i.e. whether well-studied localized paramagnetic charge carrier states, so called polaron states [4-6], either mediate spin transport by mutual interactions, or even by spatial propagation, or whether they are not involved in spin transport at all.

The involvement of charge transport mechanisms in spin-transport phenomena appears plausible, as this has been well established for inorganic semiconductors [7] and a variety of ferromagnetic (FM) resonant (FMR) spin pumping and detection studies using the inverse spin-Hall effect (ISHE) [7-11]. These studies have indicated that spin transport could be mediated through polaronic charge-carrier states [10]. However, there have also been room temperature observations of equilibrium spin diffusion in organic semiconductor materials (such as various poly-phenylene-vinylene derivates) which are known to exhibit conductivity only under non-equilibrium charge carrier injection conditions [12]. In order to explore this surprising observation, we recently conducted a comparative study between spin and charge diffusion constants for various organic semiconductors, showing that electric charge propagates orders of magnitude slower than magnetic polarization [3], implying that charge transport is not responsible for spin transport at all. This result, of course, does not exclude the possibility that free charge carrier states are nonetheless involved in spin transport, as they could possibly mediate the propagation of spin polarization via spin-spin interactions such as dipolar or exchange couplings, even when they are entirely immobile [3, 9]. However, there are also suggestions [3, 13] indicating that spin-transport does not at all involve charge carriers but rather, that it is due to spin-interactions caused by a long-range antiferromagnetic coupling.

In this Letter, we present a crucial test to conclusively demonstrate whether spin-transport in organic materials is mediated via the same electronic states as charge transport, i.e. whether it involves free charge carrier at all. By conducting a *double resonance* scheme, where a pure spin current, generated by FMR spin pumping, is subjected to electron paramagnetic resonance (EPR) induced scrambling of free charge carrier states in the organic transport layer, we aim to induce a spin current quenching, that should be observed by a reduction of the ISHE current. This is, in principle, similar to the experiments reported in Ref. 9 which, however, did not spectrally resolve the variation of the ISHE response with orientation.

In Figure 1, the concept for this experiment is shown. Panel (a) displays the control measurement: A device consisting of a FM layer adjacent to a non-magnetic organic layer whose other side interfaces a Pt film. Spin pumping of a pure spin current can be achieved through FMR excitation, and this can be confirmed by subsequent detection via the ISHE [7, 8, 11, 12, 14]. FMR excitation in the FM layer itself can be achieved using microwave (MW) radiation, and the pure spin current is then induced into the organic layer [15] via magnon-scattering at the FM/organic layer interface. The spin current through the organic layer can then be detected by a Pt layer which generates a charge current perpendicular to the spin-current via the ISHE, due to its inherent spin-orbit coupling (SOC). The current response at FMR excitation condition is therefore a direct measure of the spin current $J_S$ in the organic layer. Fig. 1(b) shows the same experimental setup with an additional, simultaneous, EPR excitation in the organic layer via MW irradiation. If the MW happens to be on-resonance with the paramagnetic charge carrier spin states in the organic layer, these states will nutate and the charge carrier ensemble will lose its spin polarization and, hence the spin current $J_S$ generated by FMR pumping, will disappear or be strongly quenched. In consequence, the measured ISHE induced charge current would disappear or be reduced, as well.

In order to carry out the experiment illustrated in Fig. 1, we fabricated pulsed ISHE device [8, 12] that use tris(8-hydroxyquinolinato) aluminum ($Alq_3$)—a small molecule that exhibits a relatively strong ISHE signal due to its increased SOC strength relative to most organic semiconductor materials, as the non-magnetic organic spin transport layer [4, 9, 16-19]. Our standard ISHE devices have ferromagnet thin films adjacent to non-magnetic active layers with appropriate contacts to detect the ISHE currents. For this study, the devices were fabricated on a glass template having a special geometry to accommodate for the design of the MW resonator cavity. First, two Al electrodes with thickness of 150 nm were deposited using photolithography. On top of the Al layers, two Cu electrodes with a separation of 50 μm were then deposited using e-beam evaporation under vacuum with a base pressure of $3 \times 10^{-7}$ torr. For the ferromagnetic layer, a 15 nm thick $Ni_{80}Fe_{20}$ thin-film was deposited using electron-beam evaporation at a rate of 0.02 nm/s. This was then followed by a 35 nm thick layer of $Alq_3$ deposited using thermal evaporation at a rate of 0.01 nm/s. For the ISHE detection, a 7 nm thick layer of Pt was then deposited also using e-beam evaporation at a rate of 0.005 nm/s. This layer serves exclusively to convert the spin current to a detectable charge current via the ISHE due to its strong SOC. After fabrication, all devices were coated with an insulating layer of $SiO_2$ (300 nm) in order to reduce anomalous-Hall effect artifacts during the ISHE measurement, as explained in Ref. 7. A grain of BDPA (a crystalline 1:1 complex of α,γ-bisdiphenylene-β-phenylallyl and benzene, Sigma-Aldrich 152560) was mounted on the device as a spin-marker [20-22] for the identification of the g ≈ 2 resonance. Continuous wave EPR and FMR as well as pulsed ISHE experiments were carried out using a commercial Bruker E580 X-band spectrometer [8, 12]. For these experiments, the devices were placed in a FlexLine ER 4118 X-MD5 dielectric resonator inside an Oxford CF935 cryostat. Continuous wave FMR and EPR were first measured using the E580's MW bridge, under continuous wave magnetic field modulation conditions, with a modulation amplitude of 0.1 mT, a frequency of 100 kHz, and a MW power of 1.966 mW. For the pulsed ISHE detection the current response was measured using a Stanford Research Systems SR570 transimpedance amplifier (using a gain of 2 μA/V and a high pass filter with a cut-off frequency of 10 Hz) whose output was recorded by the SpecJet digitizer of the E580. The MW power for the pulsed ISHE experiments was 1 kW, the pulse length was 500 ns with a duty cycle of 0.025%.

In Fig. 2 the dependence of the FMR resonance field on $\theta_B$, the angle between the magnetic field $B$ and the device plane is shown (squares; the connecting solid lines serve as guides for the eye). The angular dependence follows the well-known functionality described in Ref. 23, with the resonance field $B_0$ exhibiting a minimum for $\theta_B = 0°$ and 180° and diverging for $\theta_B = 90°$. The ISHE resonance occurs at the same resonance field $B_0$ as the FMR. The insets in Fig. 2 show color maps of transients of the ISHE current as a function of magnetic field for selected sample orientations ($\theta_B = 0°$, 72°, 108°, and 180°, indicated by red squares). Note that the current maps and the plots of the FMR resonance centers are plotted on the same magnetic field scale. The centers of both signals show good agreement. Note also that the current resonance reverses polarity between $\theta_B < 90°$ and $\theta_B > 90°$, which is characteristic of the ISHE signal [8, 11, 12]. The horizontal dashed line near 350 mT indicates the EPR resonance field for BDPA and as well as the charge-carrier spin states in $Alq_3$ whose parameters (Landé factor, hyperfine field distributions, SOC induced g-strain, spin-relaxation times), have been recently obtained with electrically detected magnetic resonance (EDMR) spectroscopy [4]; the conditions described in Fig. 1(b) can therefore be established by adjusting $\theta_B$ so that FMR and EPR overlap at $\theta_B = 72°$ or 108°.

In Fig. 3, the results of EPR and FMR measurement of the device stack, as described above, for $\theta_B = 72°$ are plotted in panels (a) and (b), together with EDMR spectra of charge carriers as obtained from bipolar injection devices similar to the measurements reported in Ref. 4, shown in (c) and (d), as well as the ISHE signal that was measured simultaneously with the EPR and FMR data in (a) and (b), as shown in (e) and (f). The left-side panels [i. e. panels (a), (c), (e), and (g)] show spectra measured over a relatively broad magnetic field range that covers the entire FMR and ISHE resonances, whereas the right-side panels [i. e. panels (b), (d), (f), and (h)] show spectra measured over a much narrower field range for clarity and with higher magnetic field resolution. Note that, as the data in panels (a) through (h) was obtained from different samples and different experiments, the microwave resonator was tuned differently for each of these. Thus, the employed magnetic resonant excitation frequencies varied slightly (by several MHz) around 9.7178 GHz, which was the actually employed frequency for the data in Fig. 3(e). In order to make the plotted data set comparable on the displayed magnetic field scale, the offset in MW frequency is considered by normalizing the spectra to a MW frequency of 9.7178 GHz for the data in panels (a) through (d), (f) through (h). The magnetic field axes of all individual spectra are therefore directly comparable. The gray shaded areas in panels (b), (d), (f), and (h) represent the excitation bandwidth of the high-power MW pulses used for the ISHE excitation, which corresponds to a magnetic field strength of ~1 mT. The MW absorption signal (blue trace) in panels (a) and (b) of Fig. 3 show contributions of the NiFe FMR (wide feature) and the BDPA EPR (narrow feature). The FMR signal from the device and the

EPR signal from the BDPA marker coincide in magnetic field, indicating that the sample orientation is chosen appropriately such that FMR and EPR resonances overlap. Panels (c) and (d) show the EDMR signals of an $Alq_3$ organic light-emitting diode (OLED) measured under comparable conditions (red trace, measured using magnetic field modulation and numerically integrated [4, 24]). This resonance originates directly from the polaron states in $Alq_3$ [4]. The EDMR resonance displays a full width at half maximum (FWHM) of 2.38 ± 0.01 mT, in good overlap with the BDPA signal since both, charge-carrier $g$-factors of $Alq_3$ and the free radical in BDPA are close to the free electron $g$-factor. The ISHE current (black open circles) along with a least-squares-fitted spectral line shape (green solid line) are shown in panels (e) and (f). In (e), a superposition of a Lorentzian line with a line width of 27.40 ± 0.06 mT and a derivative Lorentzian line (to consider a small contribution of the anomalous Hall effect, cf. Ref. [12]) is used for the numerical fit, whereas in panel (f), a third-order polynomial function was used due to the absence of a proper baseline of the resonance. Note that the experimental spectrum in Fig. 3(f) displays a much smaller step size in increments of the magnetic field, compared to Fig. 3(e). Panels (g) and (h) show the residuals, i. e. the differences between the measured ISHE data points and the fitted functions, along with their respective histograms.

From Fig. 3(e), we see that the ISHE resonances FWHM of 27.40 mT is much wider than the FWHM of 2.38 mT of the EDMR resonance associated with the polaron states in $Alq_3$ shown in Fig. 3(c) and (d). A quenching of the ISHE signal due to the EPR excitation of charge carrier and the associated resonant scrambling of charge carrier spin polarization would therefore cause an imprint of the EDMR spectrum (red trace) in the ISHE response (green trace), unless the EPR excitation were to be incomplete, i.e. only a small fraction of the charge carrier ensembles were to be in excited at any given time. In order to estimate the fraction of the charge carrier ensemble that is excited simultaneously by the high-power MW pulses used during the ISHE measurement, we convolute the EDMR resonance line function with (i) a box function, and, more realistically, (ii) with a Lorentzian function with a width of 1 mT, which corresponds to the excitation bandwidth of these pulses (i. e. the gray shaded area in Fig. 3). We thus find that (i) 35.15% or (ii) 25.79% of the entire EDMR spectrum is excited—at resonance—by the pulsed ISHE excitation. Therefore, we anticipate a quenching of the pulsed ISHE current by the same amount, i. e. a spectral hole in the ISHE spectrum. The residuals in Fig. 3 (g) and (h) are scattered around zero, and their respective histograms (red) appear approximately Gaussian distributed, with mean values of -4.15 fA [Fig. 3(g)] and -4.69 aA [Fig. 3(h)] and standard deviations of 3.05 nA and 0.544 nA, respectively. No distinct feature that would reflect the $Alq_3$ EDMR resonance can be seen. The observed quenching of the ISHE current is thus significantly smaller than the anticipated spectral hole.

Another reason why the fraction of charge carrier spins in resonance with the externally applied MW field may be smaller than anticipated arises from the magnetic stray fields generated by the FM layer in the vicinity of the spin transport layer. These random magnetic fields can potentially be superimposed upon the externally applied magnetic field and shift the EPR resonance within the $Alq_3$ layer, leading to an overall shift and widening of the EPR spectrum. We analyze this hypothesis in detail as discussed in the Suppl. Information [24] and conclude that such effects do not play a role in the experiments reported here. Thus, we conclude that magnetic resonant scrambling >25% of the charge carrier spin states took place during the experiments described above and that, due to the absence of any effect of this process on the observed ISHE currents, the spin states of free charge carriers in $Alq_3$ are not involved in spin-transport effects.

One possible explanation for the absence of any quenching of the ISHE current under EPR excitation can be that the spin-transport is mediated by paramagnetic impurities that are invisible to EPR [25]. Few models are based on the exchange mediated diffusion assumption, where spin-transport is believed to be mediated via direct exchange coupling at high carrier concentrations [1]. On the other hand, the theoretical calculations and experimental results for $Alq_3$ based spin-valves as reported in Ref. 13 asserts on spin-transport mediation via impurity band. The conclusion in ref ?? converges towards spins interacting with long range antiferromagnetic coupling, that we also support.

In summary, we have experimentally tested and rejected the hypothesis that spin-transport in the organic semiconductor $Alq_3$ is mediated through localized charge carrier (polaron) spin states. EPR induced scrambling of polaron spin does not lead to a change of an ISHE detected FMR pumped spin-current. We therefore conclude that spin-transport is not due to the propagation of charge carriers, as previously reported [3], but also not due to spin-transfer between the free, i.e. weakly exchange coupled charge carriers. We therefore attribute the ISHE induced spin-transport effects to either strongly coupled charge carriers, which, by definition, are then not free charge carriers at all, or spin-interactions caused by a long-range antiferromagnetic coupling within the electronic states $Alq_3$.

## ACKNOWLEDGMENTS

This work has been supported by the National Science Foundation, NSF-DMR #1701427. H. Malissa was supported by the U.S. Department of Energy, Office of Basic Energy Sciences, Division of Materials Sciences and Engineering under Award #DE-SC0000909.

**FIGURES**

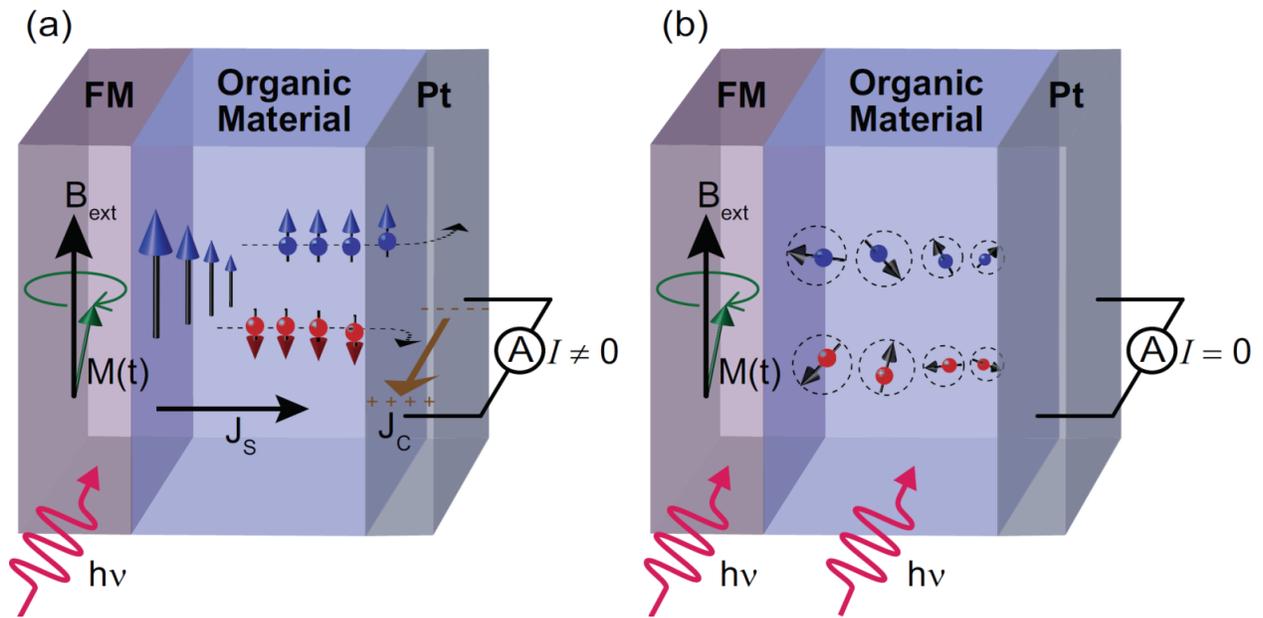

FIG. 1. Illustration of the FMR/EPR double resonance experiment reported. (a) The control experiment: Generation of a pure spin current through FMR pumping and subsequent conversion to a charge current through ISHE. (b) Experiment as in (a) with simultaneously added EPR excitation of charge carrier spin states in the organic material. The charge carrier EPR scrambles the ensemble spin polarization and, therefore, diminishes the pure spin current and the subsequent ISHE response.

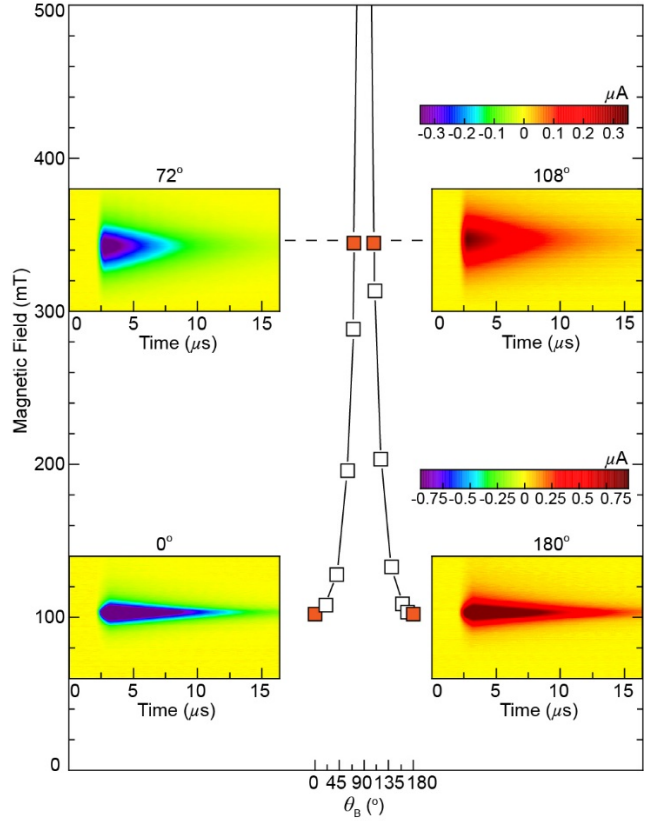

FIG. 2. Solid line: Plot of the NiFe FMR peak center as a function of the angle $\theta_B$. The color-coded data displays the ISHE current as a function of time for various $\theta_B$ (the top color bar pertains to the heatmaps corresponding to 72º and 108º, and the bottom color bar to 0º and 90º). The dashed line represents the magnetic field value that corresponds to the EPR resonance of Alq$_3$ and BDPA. The red squares represent values of $\theta_B$ which correspond to the displayed ISHE data sets.

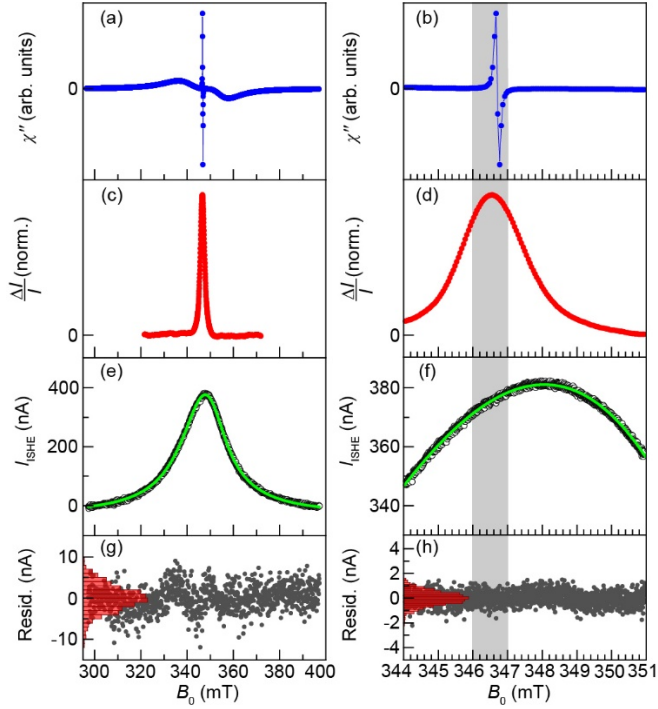

FIG. 3. (a) Plot of the X-band MW absorption measured on the ISHE sample stack. The data reveals the EPR signal of the NiFe FMR and the BDPA (cf. Ref. 20) resonance at g ≈ 2, both of which are nearly identically centered for a sample orientation of $\theta_B = 72°$. (b) Same as panel a, with a narrower magnetic field range; the gray shaded area represents the excitation bandwidth of the high-power MW pulse used in the ISHE experiment. (c) Plot of the EDMR resonance of $Alq_3$, measured on an OLED device similar to that in Ref. 4 using field modulation and subsequent numerical integration, with the resonance position normalized to MW frequency used in the ISHE measurement (cf. Refs. 4, 24). (d) Same as panel (c), measured with a narrower magnetic field range and a higher magnetic field resolution. (e) ISHE spectra (i. e. integrated current transients) measured under the same conditions as the FMR in the top panel. The green trace represents a numerical least-squares fit of a superposition of a Lorentzian function and a derivative Lorentzian function, as described I Ref. 12. (f) Same as panel (e), with a narrower magnetic field range. Here, the green trace represents a numerical least-squares fit of a third-order polynomial function due to the lack of a proper baseline. (g) Plot of the fit residuals of (c) as well as a histogram of these residuals. (h) Same as panel (g) with a narrower magnetic field range.

# Spin-transport in an organic semiconductor without free charge carrier involvement
# Supplemental Material

H. Popli,[1] J. Wang,[1] X. Liu,[1] E. Lafalce,[1] T. H. Tennahewa,[1] H. Malissa,[1] Z. V. Vardeny,[1] and C. Boehme[1]

[1]Department of Physics and Astronomy, University of Utah, Salt Lake City, Utah 84112, USA

## OLED FABRICATION

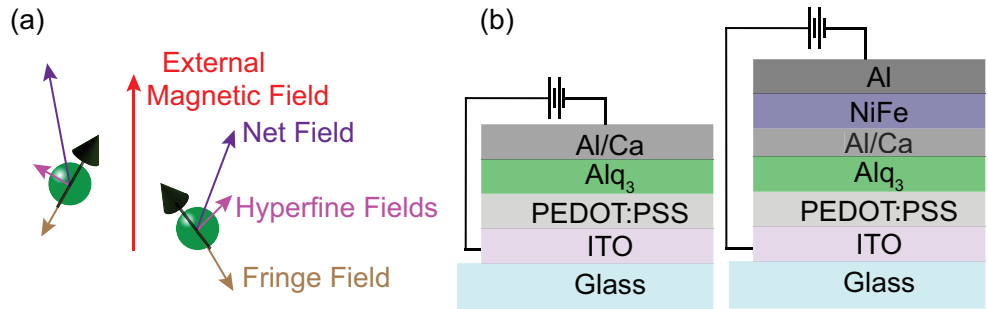

FIG. S1. (a) Schematic picture of charge carriers with their respective local magnetic fields (cf. Ref. S1). (b) Schematic of the OLED structure with (right) and without (left) adjacent ferromagnetic layer (cf. Ref. S2).

Two kinds of $Alq_3$-based OLED devices were fabricated in order to perform the control measurements to investigate the effect of fringe fields on the active layer in the ISHE device emanating from the NiFe ferromagnetic films. A pristine $Alq_3$ based OLED, i. e. without a NiFe film present and other $Alq_3$ OLEDs with varying thicknesses of the NiFe film were used. The fabrication process of the pristine devices is explained in detail elsewhere [S2]. For the OLEDs with NiFe films, the fabrication process follows the same recipe as described in the above citation until the deposition of $Alq_3$ thin films. After the $Alq_3$ layer is thermally evaporated on the substrate, a combination of extremely thin film of 20 nm of Al/Ca is thermally deposited followed by deposition of NiFe thin film and finishing off with Al as top electrode. The additional Al/Ca layer between $Alq_3$ and NiFe becomes important due the constraint in the glove box facility available as the deposition of NiFe film takes place in a separate electron-beam evaporator at a rate of 0.02 nm/s in vacuum. Several devices with varying NiFe film thicknesses, i. e. 3 nm, 15 nm, and 30 nm were used for this study.

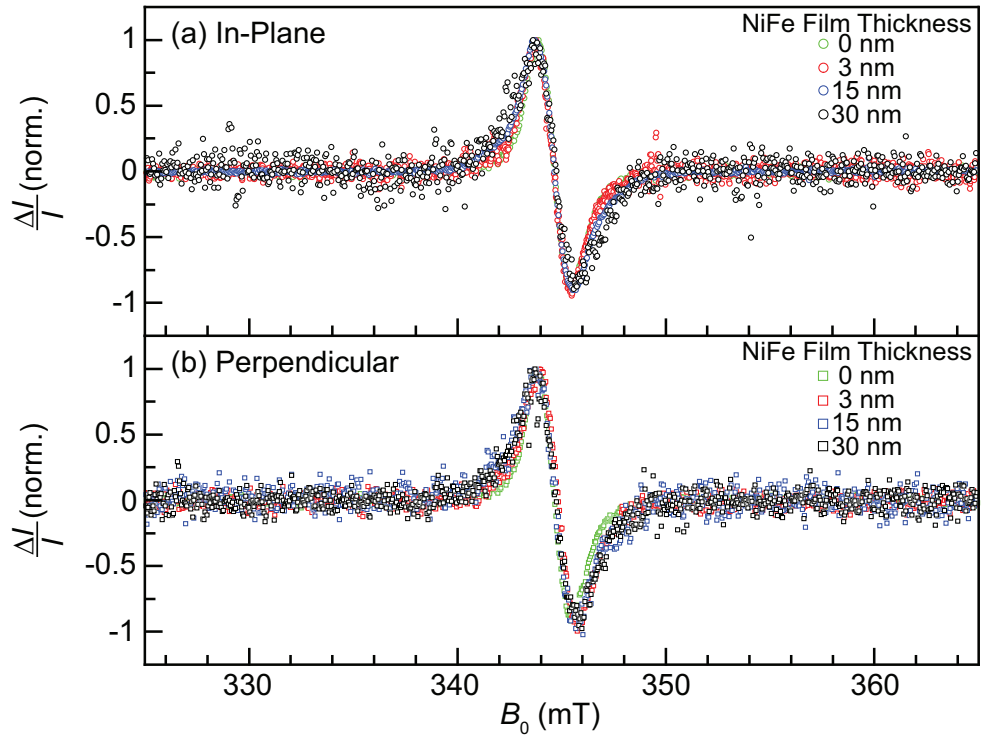

FIG. S2. Normalized EDMR spectra from Alq$_3$ OLEDs with and without an adjacent ferromagnetic layer of thicknesses ranging from 3 nm to 30 nm with the external magnetic field oriented parallel (top panel) and perpendicular (bottom panel) to the ferromagnet.

## THE EFFECT OF A FERROMAGNETIC FILM ON THE EPR IN Alq$_3$

The magnetic fields generated by a thin ferromagnetic layer [S3] may influence the EPR resonance in an adjacent Alq$_3$ film in a way that these stray fields would be superimposed on the external magnetic field, which could potentially lead to variation in the effective g-factor and line width of the individual EPR resonances. Stray field effects on magneto-resistance [S4-S6] as well as the effects fringe fields on magnetic-field effects (magneto-resistance and magneto-electro-luminescence) in OLEDs with adjacent ferromagnetic films [S1, S7-S9] have been observed previously. In order to assess the magnitude of such an effect, we compared the measured EDMR signals from OLEDs with integrated ferromagnetic layers of various thicknesses and for various orientations of the external magnetic field.

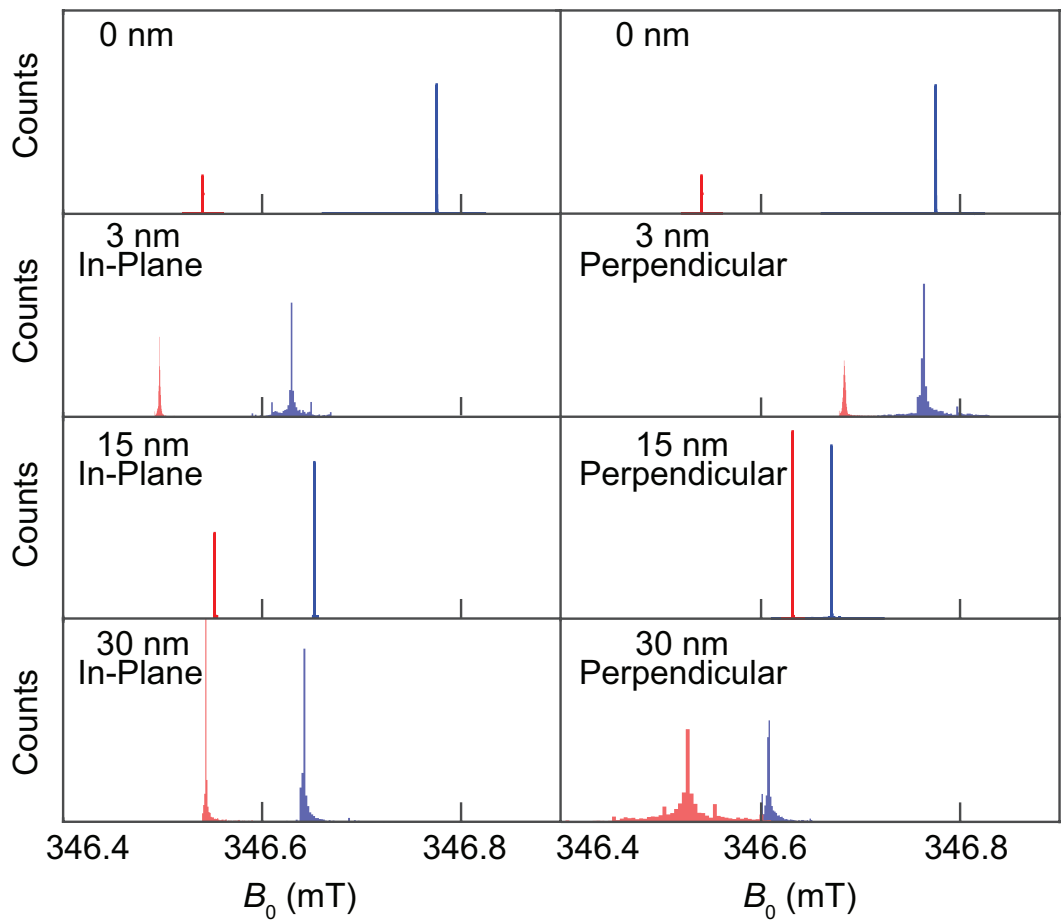

FIG. S3. Histograms of the resonance field values for both Gaussian components (cf. Ref. S2) of the EDMR lines for various thicknesses of the adjacent ferromagnetic layer at both orientations of the external magnetic field with respect to the layer.

We carried out continuous wave electrically detected magnetic resonance measurements [S2] on these various OLED devices with the external magnetic field being oriented either parallel or perpendicular to the ferromagnetic layer. These experiments were carried out using a commercial Bruker E580 X-band spectrometer at room temperature. The OLEDs are placed inside a FlexLine ER 4118X-MD5 dielectric microwave resonator in an Oxford CF935 cryostat, and a bias voltage that was adjusted between 3.6 V and 6.5 V for the various devices in order to maintain a device current of 50 µA is applied using a

Stanford Research Systems SIM928 battery source. The changes in OLED current were measured using a SR570 transimpedance amplifier with a gain of 1 µA/V for the pristine OLEDs, 200 nA/V for OLEDs with 3 nm of NiFe, and 2 µA/V for OLEDs with 15 nm and 30 nm of NiFe and a bandpass filter with a passband ranging from 100 Hz to 30 kHz for all OLEDs with NiFe, and 30 Hz to 30 kHz for the pristine OLEDs. The output signal of the SR570 is recorded using the built-in lock-in amplifier of the E580 spectrometer using field modulation.

Figure S2 shows several EDMR spectra, measured on devices with the various NiFe thicknesses for different orientations of the external magnetic field with respect to the ferromagnetic layer. The spectra were normalized to an amplitude of 1, and the magnetic field scale was adjusted to a MW frequency of 9.6649 GHz.

All spectra for the various orientations of the external magnetic field and the various NiFe thicknesses overlap and a systematic change in resonance position cannot be directly observed. In order to further corroborate the question whether a change of the resonance field occurs, we conducted a so-called *bootstrap* analysis on all measured spectra [S10-S14]: for each experimental spectra we (i) performed a least-squares fit with a double-Gaussian function with the two resonance positions and the two line widths as fit parameters and (ii) repeatedly generated hypothetical spectra by re-sampling the residuals of the fit result, and perform another least-squares fit on the artificial datasets. This procedure yields a distribution for each of the fit parameters, with a spread that reflects the noise in the original datasets.

In Figure S3 the histograms for both resonance position fit parameters (red histograms: narrow Gaussian line, blue histograms: broad Gaussian line) for the various sample orientations and NiFe thicknesses are shown. The separation of the two resonances, the variation of resonance position with the thickness of the FM layer, and the uncertainty of the fit parameters determined from the bootstrap procedure are approximately on the same order. Between the pristine OLEDs (without FM) and the OLEDs with 15 nm NiFe (as used in the ISHE experiments), the resonances on average shift by 0.0122(5) mT and 0.092(2) mT for the in-plane orientation, and 0.123(1) mT and 0.124(6) mT for the perpendicular orientation, for the red and the blue histograms, respectively.

These variations are much less than the line width of the EDMR resonances in $Alq_3$ [S2]. From this observation we conclude that the potential shifts of the EPR resonance due to the presence of an adjacent FM layer cannot account for a failure of the EPR excitation to influence the FMR-generated spin current.